\documentclass[prb,twocolumn,groupedaddress,showpacs,preprintnumbers,amsmath,amssymb]{revtex4}

\usepackage{graphicx}% Include figure files
\usepackage{dcolumn}% Align table columns on decimal point
\usepackage{bm}% bold math

\begin{document}

%%\preprint{Phys. Rev. B}

\title{Improved Light Absorption by Quantum Confinement and Band \\Folding: Enhanced Efficiency in Silicon Based Solar Cells}

\author{T. Sandu }
 \email{sandu@nanofab.uta.edu}
\author{W. P. Kirk}%
 \email{kirk@nanofab.uta.edu}
\affiliation{%TEES NanoFAB Center\\Electrical Engineering Department\\
   University of Texas at Arlington\\ Arlington, Texas 76019}%


\author{}
%\thanks{}
\affiliation{}


\date{\today}

\begin{abstract}
The improvement of light absorption in Si/BeSe$_{0.41}$Te$_{0.59}$ 
heterostructures for solar cell applications is studied theoretically. 
First, using simple approaches we found that light absorption could be 
improved in a single (uncoupled) quantum well with a thickness up to 20 
{\AA}. Second, by semiempirical tight-binding methods we calculated the 
electronic structure and optical properties of various (Si$_{2})_{n 
}$/(BeSe$_{0.41}$Te$_{0.59})_{m}$ [001] superlattices. Two bands of 
interface states were found in the band gap of bulk Si. Our calculations 
indicate that the optical edges are close to the fundamental band gap of 
bulk Si and the transitions are optically allowed.\end{abstract}
%\keywords{}

\maketitle

%\section{}
%\subsection{}
%\subsubsection{}

\section{Introduction}

\begin{figure}
\includegraphics{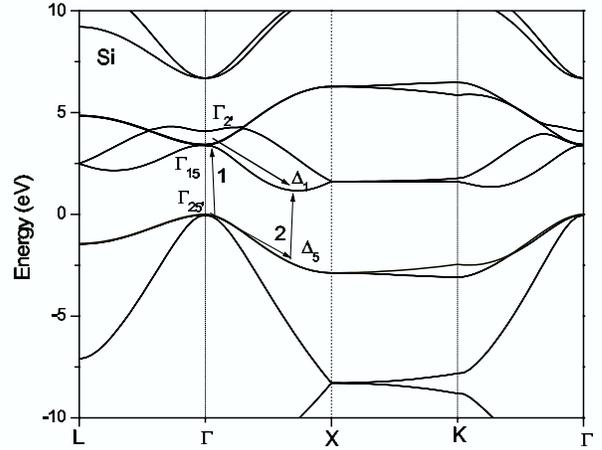}
\caption{\label{fig:1} The band structure of bulk Si with the parameters of 
\citeauthor{6} \cite{6} adapted with the spin-orbit constant \textit{$\Delta $} = 0.045 eV. The single 
group notations are used since the spin-orbit splitting cannot be distinguished 
on the chosen scale}
\end{figure}

Silicon is the principal material for integrated circuits. Not only is the 
band gap (1.12 eV) ideal for room temperature operation, but also the oxide 
(SiO$_{2})$ provides the necessary flexibility to fabricate millions of 
devices on a single chip. High integration implies high-speed operation that 
is limited by the interconnect propagation delay of the signal between 
devices. This constraint suggests that the integration of Si 
micro-electronics might be aided by optical interconnection. Unfortunately, 
silicon does not respond strongly to optical exitations because it is an 
indirect band gap semiconductor: the band extrema for electrons and holes 
are located at different points in the Brillouin zone (Fig.~\ref{fig:1}). Therefore, 
intrinsic formation or recombination of electron-hole pairs becomes a 
three-particle event, which is weaker than a two-body process. 

There has been a tremendous effort in exploring ways of breaking the silicon 
lattice symmetry and mixing different momentum states in order to induce 
optical gain \cite{1,2}. The radiative efficiency depends on the competition 
between non-radiative fast processes and relatively slow radiative 
processes. To optimize the efficiency we have to eliminate non-radiative 
channels by having high purity materials and increase the oscillator 
strength of radiative channels. The above criteria also apply to 
photovoltaics. 

One modality for improving the optical response is quantum confinement. 
Confinement of the charge density in quantum wells (QWs) permits the 
relaxation of the optical selection rules for interband transition. In 
addition, the band folding in superlattice (SL) structures will enhance the 
absorption. Consequently a Si based SL will reduce the symmetry which 
translates into band folding toward the zone center; and as a result 
vertical transitions will be available at energies closer to the indirect 
band gap. 

Beryllium-chalcogenides are good candidates for Si-based heterostructures. 
They are wide-band gap zinc blende semiconductors with lattice constants 
close to that of Si. Thus BeTe and Be Se have the lattice constants of 
5.6269 and 5.1477 {\AA}, respectively, 3.6 {\%} larger and 5.2 {\%} smaller 
than Si. Vegard's law indicates that the lattice matched composition with Si 
is BeSe$_{0.41}$Te$_{0.59}$. Recent developments \cite{3,4} in the growth of 
silicon lattice-matched BeSe$_{0.41}$Te$_{0.59}$ open the opportunity for a 
new class of Si based devices.

We analyze quantum confinement and band folding in silicon based 
heterostructures (Si/BeSeTe heterostructures) as ways of enhancing light 
absorption and, therefore, improving the efficiency of solar cells. In this 
study we use a semiempirical tight-binding (TB) method \cite{5}. In section II 
we present the tight binding parameters of BeTe and BeSe, fitted to first 
principles calculations. The model is that of \citeauthor{6} \cite{6} modified according 
to Chadi \cite{7} as to include the spin-orbit interaction. The bulk TB 
parameters of BeSe$_{0.41}$Te$_{0.59}$ are considered as virtual-crystal 
approximation of constituent materials. The TB parameters of silicon are 
those of \citeauthor{6} \cite{6} adapted to include the spin--orbit coupling. In section III we discuss some methods to improve the light absorption in Si based heterostructures such as quantum confinement in a 
single (uncoupled) QW and band folding in SL structures. A significant 
increase in direct absorption is obtained for a narrow uncoupled silicon QW. However, thin barriers are also required in order for carriers to move across the structure. In this spirit, we calculate the electronic and optical properties of Si/BeSeTe SLs using the mentioned TB method. In the last section the conclusions are outlined. 

\section{Nearest neighbor Tight-Binding parameters of beryllium chalcogenides}

Since both Si and BeSe$_{0.41}$Te$_{0.59}$ are indirect band materials, full 
band calculations are required for the theoretical understanding of electron 
transport and optical properties in such heterostructures. For this purpose 
we use the empirical tight-binding (ETB) method as one of the most used 
tools in research of complex molecular and solid-state systems \cite{8}. 
Despite the fact that ETB is based on physical approximations such as a 
one-particle picture, short-range interactions, etc. (aposteriori justified, 
however), it gives fast and satisfactory results. In terms of accuracy, it 
lies between the very accurate, very expensive \textit{ab-initio} and the fast but limited-accuracy empirical methods. TB techniques are increasingly employed 
in structures like resonant tunneling diodes, QWs, and SLs. They are more 
complete than envelope-function or \textbf{k}$ \cdot$\textbf{p} methods because they 
incorporate the entire band structure of the constituent bulk materials in a 
transparent manner. Also the short-range nature of the model is suitable for 
modeling heterointerfaces which are present in such quantum structures. In 
ETB models the electronic wave functions are expanded as a linear 
combination of atomic orbitals. The real attractiveness resides in the great 
versatility of the method that can be used for a variety of physical 
situations like: (i) molecular models based on linear combination of atomic 
orbitals with strong directional properties; (ii) calculation of electronic 
states in unknown systems based on transferability of the Hamiltonian matrix 
elements as determined in known cases (i.e. transferability of the 
Hamiltonian matrix elements from bulk semiconductors to low dimensional 
systems like QWs, quantum dots, etc.); (iii) quantitative predictions of 
physical properties (like fitting the band structure of a semiconductor to 
\textit{ab-initio} and experimental data and using those parameters to calculate optical properties of the same semiconductor). The last two points will be used in this paper. 

One of the most popular TB models is the \textit{sp}$^{3}s^{*}$ model \cite{6}. This is a twenty-band model if spin-orbit coupling is included \cite{7}. The presence of additional $s^{*}$ orbital is able to reproduce the band gap in indirect 
semiconductors like silicon. The TB parameters of silicon are those from \citeauthor{6} 
\cite{6} augmented with spin-orbit coupling according to \citeauthor{7} \cite{7}. 

BeTe and BeSe are quite new materials in the sense that there are few 
experimental facts about these semiconductors. The bulk TB parameters of 
BeTe and BeSe are determined by fitting the \textit{ab-initio} calculations of Fleszar and Hanke \cite{9}. 
Virtual-crystal approximation (VCA) is used to determine the TB parameters of BeSe$_{0.41}$Te$_{0.59}$ . 
The approximation is expected to work well because BeTe and BeSe crystallize in zinc-blende structures 
and have similar band structures. The nearest neighbor parametrization in the \textit{sp}$^{3}s^{*}$ model 
requires 15 parameters including spin-orbit interaction. The spin-orbit coupling was included due to the 
large spin-orbit splitting of the valence band, of about 0.45 eV for BeSe and 0.96 eV for BeTe. To find 
the nearest neighbor parametrization we used a similar procedure to Ref. \onlinecite{10}. The procedure is more 
consistent than that presented in Ref. \onlinecite{6}. We adapted the procedure in order to consider spin-orbit coupling. 
The results are shown in Table~\ref{tab:table1}. 

\begin{table}%[htbp]
\caption{\label{tab:table1}Matrix elements in \textit{eV} of nearest neighbor \textit{sp}$^{3}s^{*}$ model including 
spin-orbit interaction. The notation is according to \citeauthor{6} \cite{6}. Virtual-crystal approximation was used to calculate the matrix elements of 
BeSe$_{0.41}$Te$_{0.59}$. }
\begin{ruledtabular}
\begin{tabular}{cccccccc}
%{|p{66pt}|p{95pt}|p{90pt}|p{108pt}|}
%\hline
%\multicolumn{4}{|p{360pt}|}{}  \\
%\hline
& 
BeTe  \par & 
BeSe \par & 
BeSe$_{0.41}$Te$_{0.59}$ \par  \\
\hline
E(s,c)& 
5.11241& 
5.56003& 
5.295934 \\
%\hline
E(s,a)& 
-15.40059& 
-14.95297& 
-15.2171 \\
%\hline
E(p,c) & 
4.42741& 
5.02603& 
4.672844 \\
%\hline
E(p,a)& 
-0.29859& 
0.30003& 
-0.05316 \\
%\hline
E(s*,c)& 
30.16& 
21.666& 
26.67746 \\
%\hline
E(s*,a) & 
39.203& 
24.433& 
33.1473 \\
%\hline
V(s,s)& 
-3.303& 
-8.195& 
-5.30872 \\
%\hline
%V(s*,s*)& 
%0& 
%0& 
%0 \\
%\hline
V(sc,pa)& 
4.423& 
5.633& 
4.9191 \\
%\hline
V(sa,pc)& 
5.511& 
4.89& 
5.25639 \\
%\hline
V(x,x)& 
0.331& 
1.531& 
0.823 \\
%\hline
V(x,y)& 
6.362& 
6.324& 
6.34642 \\
%\hline
V(s*a,pc)& 
11.503& 
7.462& 
9.84619 \\
%\hline
V(s*c,pa)& 
3.11& 
4.572& 
3.70942 \\
%\hline
$\Delta _{a}$& 
0.97& 
0.499& 
0.77689 \\
%\hline
$\Delta _{c}$& 
0& 
0& 
0 \\
%\hline
\end{tabular}
\end{ruledtabular}
\label{tab1}
\end{table}

\begin{figure}
\includegraphics{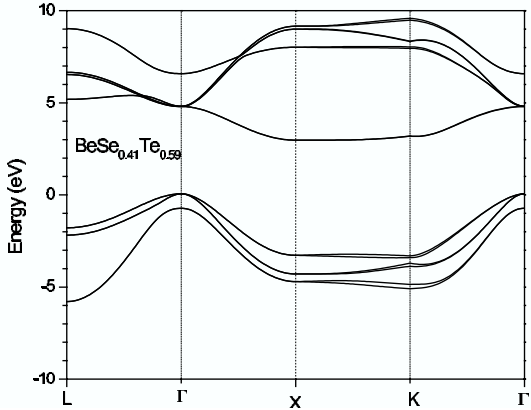}
\caption{\label{fig:2} Energy bands for BeSe$_{0.41}$Te$_{0.59}$ calculated with TB 
parameters from Table 1.}
\end{figure}

The focus of these calculations was to find the best parameters that reproduce the valence band edges ($\Gamma _{8}$ and $\Gamma _{7}$ and conduction band edge at $\Gamma _{6}$ and X$_{1})$. 
The reproduction of band edges at the zone center of 
the valence and conduction bands and the band gap between valence band and 
conduction band is within 3{\%} error. The spin-orbit coupling for Be was chosen to 
vanish, because Be is a light element and its spin-orbit coupling is 
negligible. This is also consistent with the spin-orbit splitting of 0.1 eV 
for another Be-chalcogenide, BeS with the sulfur atom, lighter than selenium 
and tellurium atoms. The energy band diagram for BeSe$_{0.41}$Te$_{0.59}$ is 
shown in Fig.~\ref{fig:2}. The material appears to be indirect band gap with the 
conduction band minimum at the $X$ point. The predicted band gap is 2.97 eV. 
Previous measurements \cite{3} indicate a conduction band offset of 1.2 eV for 
the Si/BeSe$_{0.41}$Te$_{0.59}$ heterostructure. This leaves a valence band 
offset of about 0.65 eV for the Si/BeSe$_{0.41}$Te$_{0.59}$ heterostructure. 
The TB parameters and the valence band offset will be used in the next 
section in order to calculate electronic and optical properties of the 
Si/BeSe$_{0.41}$Te$_{0.59}$ SL. 

\section{Improved Optical Absorption by Quantum Confinement and Band Folding}
\subsection{A Single Quantum Well}

The absorption edges in bulk silicon are indirect and the jumping of the 
electron from valence band (VB) to conduction band (CB) is assisted by a 
phonon (in electron transitions from VB to CB the \textbf{\textit{k}} 
momentum must be conserved). The band structure of Si is shown schematically 
in the Fig. ~\ref{fig:1}. The indirect band gap is $\Delta _{1}-\Gamma 
_{25\mbox{'}}$. The absorption process involves two steps, one is 
electron-photon interaction and the other is electron-phonon interaction. 
Basically we have two processes denoted by 1 and 2. In process 1 an electron 
is first excited to the $\Gamma _{15}$ state and then by an emission of a 
phonon the electron arrives in the $\Delta _{1}$ CB state. Similarly for 
process 2, however, in this process it starts with an emission of a phonon. 
Mathematically, the transition probability is given by Fermi's Golden Rule 
for second order perturbation

\begin{widetext}
\begin{equation}
\label{eq:1}
R_{ind} = \frac{2\pi }{\hbar }\sum\limits_{k_C ,k_V } {\left| {\sum\limits_i 
{\frac{\left\langle f \right|H_{ep} \left| i \right\rangle \left\langle i 
\right|H_{eR} \left| 0 \right\rangle }{E_{i0} - \hbar \omega }} } \right|} 
^2\delta \left( {E_C \left( {k_C } \right) - E_V \left( {k_V } \right) - 
\hbar \omega \pm \hbar \omega _p } \right),
\end{equation}
\end{widetext}

\noindent
where $\left| 0 \right\rangle $ represents the initial state, $\left| i 
\right\rangle $ the intermediate state and$\left| f \right\rangle $ the 
final state. $H_{ep}$ and $H_{eR}$ are the Hamiltonians for electron-phonon 
and electron-photon interaction, respectively. The energies $\hbar \omega $ 
and $\hbar \omega _p $are the energies for photons and phonons, 
respectively. The second sum on the right-side hand is the oscillator 
strength of the transition. For this reason, there is a long absorption tail 
between 1.12 eV and and about 3 eV that reflects the indirect nature of the 
band gap. The sharp rise in absorption with increasing photon energy 
starting around 3.2 eV (380 nm) is associated with the direct transition at 
$\Gamma $ point ($\Gamma _{25\mbox{'}} \quad  \to  \quad \Gamma _{15})$ whose 
energy is 3.4 eV (365 nm). 

In direct band gap materials light absorption in 2D systems is formally 
similar to 3D systems. Although the light absorption in indirect band gap 2D 
systems looks to be similar to 3D systems, this is not the case. The 
absorption coefficient for an uncoupled Si/ BeSe$_{0.41}$Te$_{0.59}$ QW has 
basically three components \cite{11}

\begin{widetext}
\begin{eqnarray}
\label{eq:2}
\alpha \left( \omega \right) = A\;\left[ \sum\limits_{eh,\lambda_{q}} {p_{eh}^d\delta 
\left( {E_g + E_e + E_h - \hbar \omega } \right)} + \sum\limits_{eh,\lambda_
{q}} {p_{eh}^a\,n_{\lambda_{q}} \,\delta \left( {E_g + E_e + E_h - \hbar 
\omega - \hbar \omega _{\lambda_{q}} } \right)}\right.\nonumber\\ 
 \left. { + \sum\limits_{eh,\lambda_{q}} {p_{eh} ^e\left( {n_{\lambda q} + 
1} \right)\,\,\delta \left( {E_g + E_e + E_h - \hbar \omega + \hbar \omega 
_{\lambda q} } \right)} } \right],
\end{eqnarray}
\end{widetext}

\noindent
where $A$ is a constant, $p_{eh}^{d}$ gives the direct bandgap contribution 
and $p_{eh}^{a}$ gives the phonon assisted contribution with 1-phonon 
absorption, and $p_{eh}^{e}$ is the phonon assisted contribution with 
1-phonon emission. Basically,

\begin{equation}
\label{eq:3}
p_{eh} ^d = p_{cv} I_{eh} \left( {\bm{\mbox{k}}_0 } \right),
\end{equation}

\begin{equation}
\label{eq:4}
p_{eh,\lambda \bm{q}} ^e = p_{eh,\lambda \bm{q}} ^a = p_{cv} I_{eh} 
\left( {\bm{\mbox{q}}\; - \;\bm{\mbox{k}}_0 } \right)R_\lambda, 
\end{equation}

\begin{figure}
\includegraphics{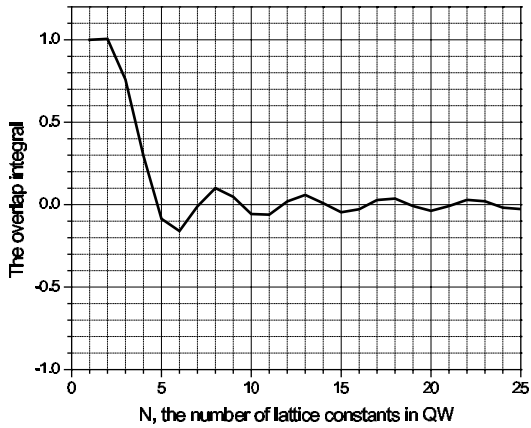}
\caption{\label{fig:3} The overlap integral (Eq.(\ref{eq:5}) for infinite wells in silicon. The 
wave vector \textbf{k}$_{0}$ in Eq.(\ref{eq:5}) is the wave vector for 
silicon conduction valleys along $z-$axis.}
\end{figure}

\noindent
$p_{cv}$ is the bulk dipole matrix element between bands, $R_{\lambda }$ is 
the matrix element contribution from electron-phonon interaction, and 
$I_{eh}$ is the overlap between the envelope-functions at different Brillouin 
points:

\begin{equation}
\label{eq:5}
I_{eh} \left( \bm{\mbox{q}} \right) = \int {d\;\bm{\mbox{r}}\;\psi _e ^{\ast} \left( 
\bm{\mbox{r}} \right)\,} \psi _h \left( \bm{\mbox{r}} \right)\;e^{ - 
i \bm{\mbox{k}}_{0}\bm{\mbox{r}}},
\end{equation}

\noindent
where $\bm{\mbox{k}}$$_{0}$ is the location in $k-$space of each conduction 
valley. For silicon, if $z$ is the growth direction, then the 2 valleys along 
z-directions (with $k_{y} = k_{x} = 0 )$ are responsible for direct transitions. The 
other four valleys contribute to phonon-assisted absorption. For infinite 
wells the overlap integral $I_{eh}$ has the behavior depicted in Fig. ~\ref{fig:3}. For 
very narrow QW's the overlap integral tends to one, which means that in a 
genuine 2D system only the transverse momentum has to be conserved. For the 
other asymptotic limit, i.e. very wide well, the overlap integral vanishes. 
In physical terms, this says that the system became genuinely 3D and any 
electromagnetic transition has to be vertical in the absence of phonons. 
Moreover, from Fig. ~\ref{fig:3} we may expect to have a strong direct transition (with 
an overlap integral no less than 0.1) for a QW with a width up to 20 {\AA}, 
i.e. an ultrathin QW. In the following we will discuss the possibility of 
improving the oscillator strength in Si/ BeSe$_{0.41}$Te$_{0.59}$ based SLs. 

\subsection{Electronic Structures of \textbf{(Si}$_{2}$\textbf{)}$_{n}$\textbf{/(BeSe}$_{0.41}$\textbf{Te}$_{0.59}$\textbf{)}$_{m}$\textbf{ [001]}
Superlattices}

The use of SL structures relaxes the condition of ultrathin QW. However, the 
SL has to have thin barriers. Thin barriers allow photo-generated carriers 
to be swept away by the built in electric field without recombination. 
Thicker barriers induce localization of carriers and implicitly higher 
recombination rate. The alternation of QWs and barriers along [001] growth 
axis will generate a band folding in the SL and band mixing of zone-center 
and zone edge states \cite{12,13}. The band mixing will inherently enhance the 
oscillator strength for direct transition in Si structures. On the other 
hand, band folding will induce states to which vertical transitions are 
possible at energies by far lower than 3.4 eV, the lowest energy for 
vertical transitions in bulk Si. 

\begin{figure*}
\includegraphics{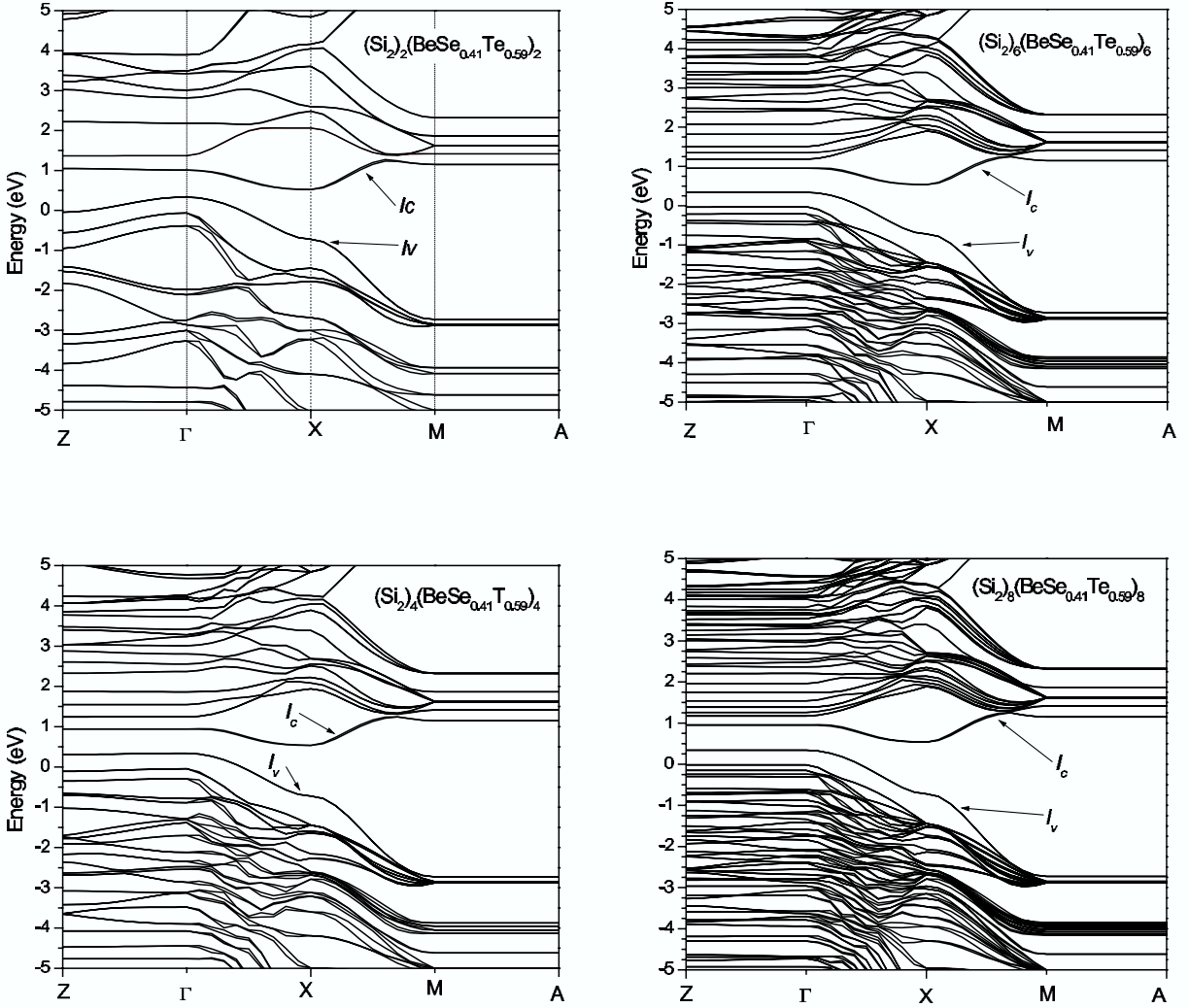}
\caption{\label{fig:4} Band structures of (Si$_{2})_{n}$/(BeSe$_{0.41}$Te$_{0.59})_{m}$ 
[001] superlattices for $n = m = 2,4,6,8$. The interface bands are \textit{Iv} and 
\textit{Ic}.}
\end{figure*}

\begin{figure*}
\includegraphics{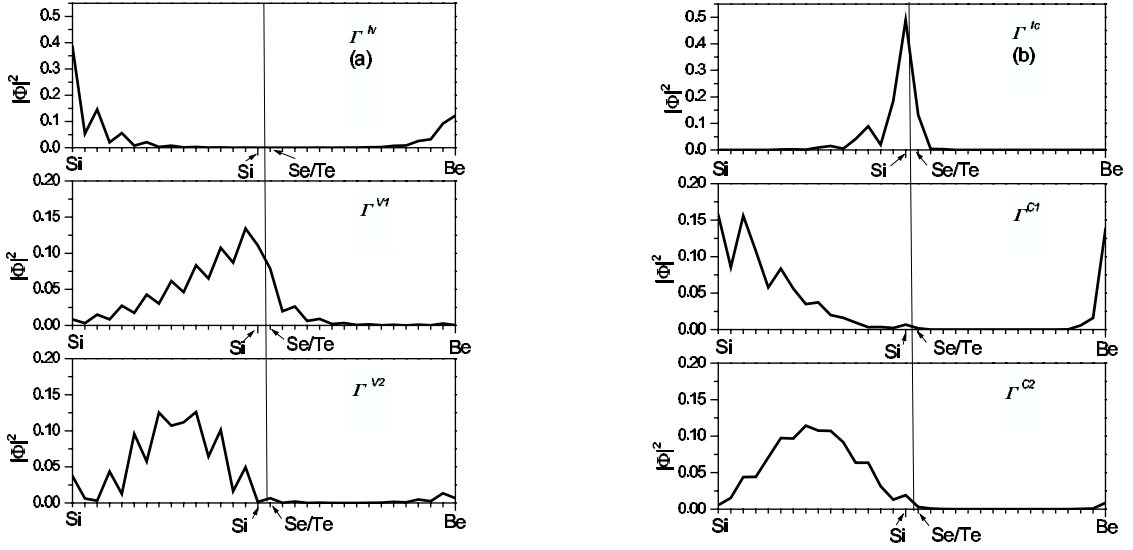}
\caption{\label{fig:5}Planar charge densities in the 
(Si$_{2})_{8}$/(BeSe$_{0.41}$Te$_{0.59})_{8}$ [001] superlattice at 
\textit{$\Gamma $} point. (a) Occupied states: the interface state $\Gamma ^{\,I_V }$, and the 
first top two confined states ($\Gamma ^{\,V1}$ and $\Gamma ^{\,V2})$. (b) 
Empty states: the interface state $\Gamma ^{\,I_c }$, and the first two 
confined states ($\Gamma ^{\,C1}$ and $\Gamma ^{\,C2})$. The solid vertical 
lines denote the interfaces.}
\end{figure*}

We consider a Si/BeSe$_{0.41}$Te$_{0.59}$ SL whose layers are perpendicular 
to [001] direction. We employ the nearest neighbor \textit{sp}$^{3}s^{*}$ TB Hamiltonian 
including spin-orbit interaction. We denote this SL as a (Si$_{2})_{n 
}$/(BeSe$_{0.41}$Te$_{0.59})_{m}$ SL with $n$ two-atom thick layers of Si and 
$m$ two-atom thick layers of BeSe$_{0.41}$Te$_{0.59}$ repeated periodically. 
The defined supercell consists of 2 ($n+m)$ adjacently bonded atoms Si, Si, 
\ldots ..Si, Be, Te/Se, Be, Te/Se, \ldots ..Be, Te/Se. The TB matrix 
elements of the Si/BeSe$_{0.41}$Te$_{0.59}$ SL are taken over directly from 
those bulk values. The on-site energies of the BeSe$_{0.41}$Te$_{0.59}$ are 
accordingly changed to match the valence band offset at the interface. 
Simple averages were used to supply the parameters connecting different 
materials at the interface. Since the spin orbit must be included the SL 
Hamiltonian will be represented as having $20\,\left( {n + m} \right)$ 
functions. Once the TB matrix elements have been established the SL band 
structure reduces to the diagonalization of the $20\,\left( {n + m} 
\right)\times 20\,\left( {n + m} \right)$ matrix Hamiltonian. The band 
structure of (Si$_{2})_{n }$/(BeSe$_{0.41}$Te$_{0.59})_{m}$ SL for $m = 
n = 2$, $m = n = 4$, $m = n = 6$, and $m = n = 8$ are displayed in Fig. ~\ref{fig:4}. 
The zero of energy corresponds to the top of the valence band in bulk Si. 
The band folding effect can be seen as many crowded subbands. Two interface 
bands ($I_{v}$ and $I_{c})$, one empty and one occupied, were found. They lie 
in the lower and upper parts of the band gap of bulk silicon, respectively. 
The origin of these interface bands rests on the polar nature of the 
interface as was also found in GaAs/Ge SL \cite{14}. The polarity of the 
interface originates from the large differences in the on-site energies for 
the constituent atoms (Si and Be or Se/Te). Even if a (110) non-polar 
interface is used, one interface band is still found in II-VI/IV SLs \cite{15,16}. 
We calculated the planar charge density of some of the occupied and empty 
band edge states of the SL with $\mbox{m} = \mbox{n} = \mbox{8}$ at $\Gamma 
$ point in the Brillouin zone. The planar charge density of some of the band 
edge states and interface states are depicted in Fig. ~\ref{fig:5} for the SL with 
$\mbox{m} = \mbox{n} = \mbox{8}$. By interface states we mean states which 
die away within few layers from the interface. We denote by $\Gamma ^{I\,v}$, 
the interface states of the $I_{v}$ band and by $\Gamma ^{I\,c}$, the 
interface states of the $I_{c}$ band at $\Gamma $ point. The charge density 
of the $\Gamma ^{I\,v}$ interface state has a maximum at the Be-Si 
interface, while the next occupied state is confined, but still is localized 
toward the Si-Se/Te interface. Only the third occupied state is genuinely 
confined in silicon slab. The charge density of the $\Gamma ^{I\,c}$ 
interface state has a maximum at the Si-Se/Te interface, while the next 
empty state is confined, but still is localized toward the Si-Be interface. 
Again, only the third empty state is genuinely confined in silicon slab. We 
believe that this is associated with acceptor behavior of Be and donor 
behavior of Se and Te with respect to Si \cite{17}. 

\subsection{ Optical Properties of \textbf{(Si}$_{2}$\textbf{)}$_{n}$\textbf{/(BeSe}$_{0.41}$\textbf{Te}$_{0.59}$\textbf{)}$_{m}$\textbf{ [001]}
Superlattices}

The electronic contribution to absorption spectrum is given by $\sigma 
_{abs} \left( \omega \right)\sim \omega \,\varepsilon _2 \left( \omega 
\right)$, where$\sigma _{abs} \left( \omega \right)$ is the absorption 
coefficient and \cite{8}

\begin{equation}
\label{eq:6}
\varepsilon _2 \left( \omega \right) = \frac{2\pi ^2\hbar \,e^2}{m\omega 
\,\Omega }\sum\limits_{c,v,k} {f_{cv,k} \delta \left( {E_{c,k} - E_{v,k} - 
\hbar \omega } \right)} 
\end{equation}

\noindent
is the imaginary part of dielectric function. Here $m$ is the electron mass, 
\textit{$\Omega $} is the volume, $e$ is the electron charge, $\hbar $ is the Planck constant and 
$f_{cv,k} $is the oscillator strength for the direct transition from the 
state $\left| {v,{\rm {\textbf k}}} \right\rangle $ to $\left| {c,{\rm {\textbf k}}} 
\right\rangle $, with the photon momentum neglected. The oscillator strength 
is defined as

\begin{equation}
\label{eq:7}
f_{cv,k} = \frac{2}{m}\frac{\left| {\left\langle {c,{\rm {\textbf k}}} 
\right|\,{\rm {\bm \varepsilon }} \cdot {\rm {\textbf p}}\,\left| {v,{\rm {\textbf 
k}}} \right\rangle } \right|^2}{E_{c,k} - E_{v,k} }
\end{equation}

In Eq. (\ref{eq:7}) $\left| {v,{\rm {\textbf k}}} \right\rangle $ and $\left| {c,{\rm 
{\textbf k}}} \right\rangle $ are the valence and conduction band, eigenstates, 
$E_{v,k}$ and $E_{c,k}$ are their corresponding energies, ${\rm {\bm 
\varepsilon }}$ is the polarization of light, and \textbf{p} is the 
momentum operator. In the empirical tight-binding approach the momentum 
matrix element is defined as \cite{18}

\begin{equation}
\label{eq:8}
\left\langle {c,{\rm {\textbf k}}} \right|{\rm {\textbf p}}\left| {c,{\rm {\textbf k}}} 
\right\rangle = \frac{m}{\hbar }\left\langle {c,{\rm {\textbf k}}} 
\right|\,\nabla _{\rm {\textbf k}} H\left( {\rm {\textbf k}} \right)\,\left| {v,{\rm 
{\textbf k}}} \right\rangle 
\end{equation}

\begin{figure}[hb!]
\includegraphics{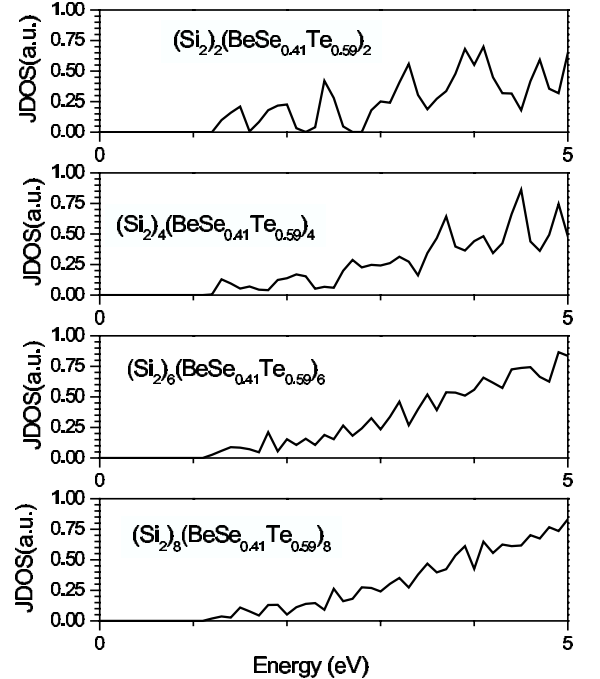}
\caption{\label{fig:6} Joint density of states corresponding to optical transitions for 
(Si$_{2})_{n}$/(BeSe$_{0.41}$Te$_{0.59})_{m}$ [001] superlattices with $n 
= m = 2,4,6,8$.}
\end{figure}

Because the variation of the oscillator strength over the Brillouin zone is 
small \cite{13}, we first calculate the joint densities of states (JDOS). We 
assume that light is propagating along the SL growth direction. The JDOS 
represents the number of states that can undergo energy and 
\textbf{k}-conserving transitions for photon frequencies between 
\textit{$\omega $} and $\omega  + d\omega $. The JDOS associated with 
Eq. (\ref{eq:6}) are shown in Fig. ~\ref{fig:6} for SLs with $m = n = 2$, $m = n = 4$, $m = n = 6$, 
and $m = n = 8$. A 0.05 eV broadening was considered for each electronic energy. 
The summation over Brillouin zone was replaced by the summation over special points 
in the Brillouin zone \cite{19,20}. 

\begin{figure}[t!]
\includegraphics{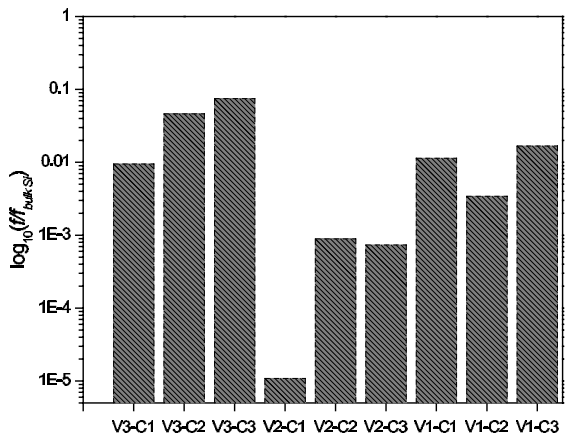}
\caption{\label{fig:7} The oscillator strengths $f$ of several interband transitions relative 
to the direct transition in bulk Si ($\Gamma _{25\mbox{'}}  \quad  \to $ 
$\Gamma _{15}) \quad f_{bulkSi}$ for the 
(Si$_{2})_{4}$/(BeSe$_{0.41}$Te$_{0.59})_{4}$ [001] superlattice.}
\end{figure}

Due to band folding the absorption edges for vertical 
transitions are lowered toward the indirect band gap of bulk Si. Moreover, 
the curves rise slowly and, with increasing $m\left( { = n} \right)$, the 
absorption edges extend to lower energies. Similar results were found in the 
calculations for porous Si with periodic boundary conditions \cite{21}. 
Eq. (\ref{eq:6}) tells us that the strength of the optical absorption is also 
determined by the oscillator strength. We denote by V1, V2, and V3 the first 
three top valence subbands, and by C1, C2, and C3 the first three conduction 
subbands. We calculated the oscillator strengths of several interband 
transitions relative to the oscillator strength of the direct transition in 
bulk Si. The results are shown in Fig. ~\ref{fig:7} for $m = n = 4$. The oscillator 
strengths of interband transitions are at least 10 times smaller than their 
bulk counterpart and basically range from 10 $^{\mbox{--}3}$ to 10 $^{ - 1}$ 
relative to $\Gamma _{25\mbox{'}} \quad  \to  \quad \Gamma _{15}$ transition in 
bulk Si. The strongest transitions are those coming from the third confined 
hole level to the second and third confined electron levels. 

\section{Conclusions}
We studied light absorption of silicon-based heterostructures 
(Si/BeSe$_{0.41}$Te$_{0.59})$ that might be used in solar cell applications. 
We found that significant absorption occurs for thin uncoupled quantum wells 
with their width up to 20 {\AA}. We also studied the 
(Si)$_{n}$/(BeSe$_{0.41}$Te$_{0.59})_{ m}$ superlattice structures with a 
semiempirical tight-binding method. In order to perform the superlattice 
calculations, we determined the tight-binding parameters for BeSe and BeTe 
in the \textit{sp}$^{3}s^{*}$ nearest neighbor model including spin-orbit interaction. 
Electronic structure and optical properties were calculated for various 
superlattice structures. Two interface bands were found to exist in the band 
gap of bulk silicon. We found that band folding induces vertical transitions 
near the indirect band gap of bulk Si. In addition, calculated oscillator 
strengths for vertical transitions near the optical band edge show the 
mixing of the zone-center and zone edge states of bulk Si for conduction 
subbands. Therefore, the transitions are optically allowed and the response 
of silicon based heterostructures to illumination is enhanced.

\begin{acknowledgments}
This work was supported in part by NASA grant NCC3-516 and by the Texas 
Advanced Technology program under grant No. 003594-00326-1999.
\end{acknowledgments}

%\appendix
%\section{}

%\begin{acknowledgments}
%\end{acknowledgments}

\bibliography{Sandu-Kirk}

\end{document}